\documentclass[12pt]{article}

\usepackage{latexsym}

\font\tenmsbm=msbm10 scaled 1200
\font\sevenmsbm=msbm9
\newfam\msbmfam
\textfont\msbmfam=\tenmsbm \scriptfont\msbmfam=\sevenmsbm

\makeatletter
\@addtoreset{equation}{section}
\makeatother

\newcommand{\eref}[1]{(\ref{#1})}

\def\be{\begin{equation}}
\def\ee{\end{equation}}
\def\ba{\begin{eqnarray}}
\def\ea{\end{eqnarray}}
\def\bet{\begin{tabular}}
\def\eet{\end{tabular}}
\def\pa{\partial}

\def\nn{\nonumber}
\def\ve{\varepsilon}

\def\a{\alpha}
\def\b{\beta}

\def\d{\delta}

\def\r{\rho}
\def\s{\sigma}
\def\m{\mu}
\def\n{\nu}
\def\ra{\rightarrow}
\def\vp{\varphi}

\catcode`@=11

\begin{document}

\begin{titlepage}

\begin{flushright}
Preprint DFPD 06/TH/06\\
June 2006\\
\end{flushright}

\vspace{2.5truecm}

\begin{center}

{\Large \bf  Radiation reaction and four--momentum conservation}
\vskip0.3truecm
 {\Large \bf for point--like dyons }

\vspace{2.0cm}

K. Lechner\footnote{kurt.lechner@pd.infn.it}

\vspace{2cm}

 {
\it Dipartimento di Fisica, Universit\`a degli Studi di Padova,

\smallskip

and

\smallskip

Istituto Nazionale di Fisica Nucleare, Sezione di Padova,

Via F. Marzolo, 8, 35131 Padova, Italia
}

\vspace{2.5cm}

\begin{abstract}

We construct for a system of point--like dyons a conserved
energy--momentum tensor entailing finite momentum integrals, that
takes the radiation reaction into account.

\vspace{0.5cm}

\end{abstract}

\end{center}
\end{titlepage}

\newpage

\baselineskip 8 mm


\section{Introduction}

A charged point--particle creates an electromagnetic field strength,
the Lienard--Wiechert field, that diverges at the position of the
particle. This implies that the radiation reaction, or self force,
experienced by the particle is infinite, and the Lorentz--equation
is ill defined. The divergent contribution from this equation can be
eliminated by a classical infinite mass renormalization, to obtain a
finite effective equation of motion, the Lorentz--Dirac equation
\cite{dirac}, that plays a crucial role in classical radiation
theory. This equation takes the radiation reaction into account and
completes the Larmor formula.

The Lorentz--Dirac effective equation of motion bears several
unusual features. It is of third order in the time derivative, and
it can not be deduced from a standard Lagrangian: eventually it must
be postulated. The ultimate justification for the equation comes
from the requirement of {\it local energy--momentum conservation},
i.e. there should exist an energy--momentum tensor $T^{\m\n}$ that
is  a) conserved, and b) admits finite momentum integrals. The {\it
naive} energy--momentum tensor $\Theta^{\m\n}$, being quadratic in
the field strength $F^{\m\n}$, diverges at the position of the
particle like $\Theta^{\m\n}\sim 1/R^4$, if $R$ is the distance from
the particle, and it does therefore not admit finite four--momentum
integrals. Said differently, while $F^{\m\n}$ is a distribution
$\Theta^{\m\n}$ is not a distribution, because the square of a
distribution is in general not a distribution. Consequently, the
four--divergence $\pa_\m \Theta^{\m\n}$ does not even make sense.

The construction of a consistent energy--momentum tensor requires,
in some sense, to isolate and subtract from $\Theta^{\m\n}$ the
singularity present at the position of the particle, {\it without
modifying the value of $\Theta^{\m\n}$ in the complement of the
particle's worldline}, in compatibility with energy--momentum
conservation and Lorentz invariance. More precisely, the so obtained
renormalized energy--momentum tensor should be conserved, {\it if
the particle satisfies the Lorentz--Dirac equation of motion}. It is
clear that such a program can be carried out only in the framework
of distribution theory.

Until now the construction outlined here has been realized only for
a charged point particle in four dimensions, rather recently
\cite{rowese}, using a somewhat cumbersome and implicit distribution
technique. A physically more transparent alternative representation
of the resulting energy--momentum tensor  -- again in the framework
of distribution theory -- has been given in \cite{LM}, relying on a
new Lorentz--invariant regularization scheme, followed by a
classical renormalization.

Aim of the present letter is to generalize the new approach of
\cite{LM} to construct a consistent energy--momentum tensor for a
system of point like dyons, taking the radiation reaction into
account, which has not been given before \footnote{The
energy--momentum tensor proposed in \cite{rohrdy} requires to modify
the naive tensor $\Theta^{\m\n}$ {\it also} in the complement of the
worldline.}. This result completes the consistency proof for a
classical system of radiating dyons, satisfying generalized duality
invariant Lorentz--Dirac equations \cite{rohrdy}, \cite{vanmeter},
see \eref{lddyon}.

The new method illustrated here, due to its manifest
Lorentz--invariance at each step, appears in particular suitable for
extension to a system of strings or branes in arbitrary dimensions.
Using this method we hope indeed to furnish elsewhere the
construction of a consistent energy--momentum tensor for a generic
radiating extendend object, that is still unknown.

\section{Regularized Maxwell--equations}

For simplicity we consider a single dyon with mass $m$, electric and
magnetic charges $e$ and $g$, and worldline $y^\m(s)$, the extension
of our construction to a system of $N$ dyons being straightforward.
We denote four--velocity, four--acceleration and derivative of the
four--acceleration by $u^\m=dy^\m/ds$, $w^\m=du^\m/ds$,
$b^\m=dw^\m/ds$. Introducing a current with unit charge as,
 \be
j^\m (x) = \int u^\m\, \d^4 (x - y)\,ds,
 \ee
the electric and magnetic currents are $j^\m_e=ej^\m$ and
$j^\m_m=gj^\m$. The Maxwell equations for the dyon become then,
 \ba
\pa_\m F^{\m\n}&=& j^\n_e,\label{mod1}\\
\pa_\m\widetilde F^{\m\n}&=&j^\n_m,\label{mod2}
 \ea
where we indicate the dual of an antisymmetric tensor with
$\widetilde F^{\m\n}\equiv{1\over 2}\,\ve^{\m\n\r\s}\,F_{\r\s}$. The
general solution of \eref{mod1}, \eref{mod2} can be written as,
 \be
F^{\m\n}=f^{\m\n}+e\, H^{\m\n}-g\, \widetilde H^{\m\n},\quad
H^{\m\n}=\pa^\m A^\n -\pa^\n A^\m,
 \ee
where $f^{\m\n}$ is a free radiation field, $\pa_\m
f^{\m\n}=0=\pa_\m\widetilde f^{\m\n}$, and $A^\m$ is a unit
Lienard--Wiechert potential in Lorentz--gauge, satisfying $\Box
A^\m=j^\m$, $\pa_\m A^\m=0$,
 \be\label{LW}
   A^\m={u^\m\over 4\pi (uR)}.
 \ee
We write the scalar products as $a^\m b_\m=(ab)$, and we have
defined,
 \be\label{retprop}
 R^\m(x)\equiv x^\m-y^\m(s).
 \ee
The proper time appearing in $y^\m$ and in $u^\m$ is the retarded
proper time $s(x)$ determined from,
 \be
 (x-y(s))^2=0, \quad x^0>y^0(s).
 \ee
This means in particular that we have $R^\m R_\m=0$, and hence
$R^0=|\vec R|\equiv R$.

Since $A^\m$ carries $1/R$ singularities, the unit field strength
$H^{\m\n}$ carries $1/R^2$ singularities near the worldline, and the
Lorentz equation for the dyon would be singular. After subtraction
of the singularity one postulates the following finite duality
invariant generalization of the Lorentz--Dirac equation for a dyon
\cite{vanmeter}, $p^\m=m u^\m$, $w^2=w^\m w_\m$,
 \be\label{lddyon}
{dp^\m\over ds}= {e^2 +g^2\over 6 \pi} \left({dw^\m\over
ds}+w^2\,u^\m\right)
 + \left(e f^{\m\n}+g\widetilde f^{\m\n}\right)u_\n,
 \ee
that takes the radiation reaction into account. For $g=0$ one gets
back the Lorentz--Dirac equation.

On the other hand, while $A$ and $F$ have at most $1/R^2$
three--space integrable singularities near the worldline of the
particle and are distributions, the naive electromagnetic
energy--momentum tensor \footnote{In the following we use for the
contraction of two antisymmetric tensors the notation
$(AB)^{\m\n}=A^\m{}_\r B^{\r\n}$, and $(AB)=A^{\m\n} B_{\n\m}$.},
 \be\label{naiv}
 \Theta^{\m\n}=(FF)^{\m\n}-{1\over 4}\,\eta^{\m\n}(FF),
 \ee
carries three--space {\it non integrable} $1/R^4$ singularities, and
is not a distribution.

As first step to isolate the singularities in $\Theta^{\m\n}$ we
introduce a Lorentz--invariant regularization, parametrized by a
positive regulator with the dimension of length $\ve$, by replacing
the retarded proper time $s(x)$ appearing in $A^\m$ in \eref{LW},
with a regularized retarded proper time $s_\ve(x)$, determined from,
 \be\label{retreg}
(x-y(s))^2=\ve^2, \quad x^0>y^0(s).
 \ee
We call the resulting regularized potential,
 \be\label{potreg}
A_\ve^\m=\left.{u^\m\over 4\pi (uR)}\right|_{s=s_\ve(x)}, \quad
\pa_\m A^\m_\ve=0,
 \ee
where from now on with $y^\m$, $u^\m$, $R^\m$ etc. we intend their
regularized versions, obtained through the replacement $s(x)\ra
s_\ve (x)$. The regularized field strength becomes then,
 \be\label{regfs}
F^{\m\n}_\ve=f^{\m\n}+e\, H^{\m\n}_\ve-g\, \widetilde H^{\m\n}_\ve,
 \ee
where,
 \be\label{regh}
 H^{\m\n}_\ve=\pa^\m A^\n_\ve -\pa^\n A^\m_\ve.
 \ee
We define the regularized energy--momentum tensor as,
 \be \label{regtmn}
 \Theta^{\m\n}_\ve=(F_\ve F_\ve)^{\m\n}-{1\over 4}\,\eta^{\m\n}(F_\ve F_\ve).
 \ee
The fields $A_\ve$, $F_\ve$ and $\Theta_\ve$ are now all regular on
the particle's worldline, indeed they are $C^\infty$--functions on
${\bf R}^4$. But, whereas $A_\ve$ and $F_\ve$ for $\ve\ra 0$ tend to
$A$ and $F$ in the distributional sense, $\Theta_\ve$ converges to
$\Theta$ {\it pointwise away from the worldline}, but not in the
distributional sense, because $\Theta$ is not a distribution
\footnote{Saying that a set of functions, or distributions, $f_\ve$
converges for $\ve\ra 0$ to $f$ in the distributional sense means
that it converges if applied to an arbitrary test function, i.e. one
has $\lim_{\ve\ra 0} f_\ve(\vp)=f(\vp)$ for every $\vp\in {\cal
S}({\bf R}^4)$.}.

\section{Construction of a finite energy--momentum tensor}

Before taking the distributional limit of  $\Theta_\ve$ we must
therefore separate and subtract its contributions that diverge for
$\ve\ra 0$ in the distributional sense. This means that we have to
apply  $\Theta_\ve$ to a generic test function $\vp$, and isolate
the terms of the resulting integral that diverge for $\ve\ra 0$. The
rest of this section is mainly devoted to the explicit
identification, and subtraction, of these singular terms, the result
being formula \eref{trenorm}.

To begin with we need an explicit expression for $F_\ve$, i.e. for
$H_\ve$. Differentiating \eref{retreg} to derive $\pa_\m s_\ve=
R_\m/(uR)$, from \eref{potreg} one obtains,
 \be\label{derpot}
\pa^\m A_\ve^\n ={1\over 4\pi(uR)^3}\left[(\eta^{\m\r}-u^\m
u^\r)R_\r u^\n+(u^\r w^\n- u^\n w^\r)R_\r R^\m\right].
 \ee
The most singular terms in $H_\ve$, see \eref{regh}, go therefore as
$1/R^2$ near the worldline, and since the radiation field $f^{\m\n}$
is supposed to be regular, the contributions of $\Theta_\ve$ that do
not converge for $\ve \ra 0$ in the distributional sense, are only
the ones quadratic in $H_\ve$. Inserting \eref{regfs} in
\eref{regtmn} and keeping only the terms quadratic in $H_\ve$, one
obtains for the divergent part of $\Theta_\ve$ therefore,
 \be\label{thediv}
\left.\Theta^{\m\n}_\ve\right|_{div}= (e^2+g^2)\left[\left(H_\ve
H_\ve\right)^{\m\n}  -{1\over 4}\,\eta^{\m\n}(H_\ve H_\ve)
\right]_{div}.
 \ee
Notice that the cross terms in $e$ and $g$ canceled.

We remain then with the evaluation of $(H_\ve H_\ve)_{div}$. This
product contains terms that behave as $1/R^n$ near the worldline,
with $n=2,3,4$. As $\ve \ra 0$ in the distributional sense, for
dimensional reasons the terms with $n=4$ give rise to simple pole
$[\sim 1/\ve]$, and logarithmic $[\sim \ln \ve]$ singularities, the
ones with $n=3$ give rise to logarithmic singularities, while the
ones with $n=2$ are convergent. Actually, it can be seen that the
logarithmic singularities cancel between the $n=4$ and $n=3$ terms,
and one remains only with the pole singularities contained in the
$1/R^4$ terms. To determine $(H_\ve H_\ve)_{div}$ it is then
sufficient to evaluate the simple pole term of, see \eref{derpot},
 \be\label{polar}
\left(\pa^\m A^\n_\ve \pa^\a A^\b_\ve\right)_{div}=\left.{1\over 16
\pi^2} \left(\eta^{\m\r}-u^\m u^\r\right)\left(\eta^{\a\s}-u^\a
u^\s\right) u^\n u^\b {R_\r R_\s\over (uR)^6}\right|_{1/\ve},
 \ee
where we kept only the $1/R^4$--terms. The rest of this section is
devoted to the explicit evaluation of the r.h.s of this formula.

Since the divergences as $\ve\ra 0$ are intended in the
distributional sense, we must apply the r.h.s of \eref{polar} to a
test function. Omitting for simplicity of writing the (regular)
tensorial prefactor $(\eta^{\m\r}-u^\m u^\r)\cdots$ in \eref{polar},
we have to evaluate the function $R_\r R_\s/(uR)^6$ applied to a
generic test function $\vp(x)\in {\cal S}$, i.e. by definition,
 \be\label{poldiv}
{R_\r R_\s\over (uR)^6}(\vp)\equiv \int {d^4x\over (uR)^6 }\,R_\r
R_\s\, \vp(x) =\int ds \int {d^4x\over (ux)^5}\, 2\,\d(x^2-\ve^2)
x_\r x_\s\, \vp(x+y),
 \ee
where we have inserted a $\d$--function to take the constraint
\eref{retreg} into account, and we have performed the shift $x^\m\ra
x^\m+y^\m$. The $x$--integration is restricted to $x^0>0$. The
kinematical quantities $y$ and $u$ are now evaluated at $s$, that is
a free integration variable. The pole singularity of this expression
can be extracted by rescaling $x^\m\ra \ve x^\m$, and sending in the
integral $\ve$ to zero \footnote{The integral \eref{poldiv} contains
actually also logarithmic divergences $\sim \ln \ve$, but these can
be seen to cancel against the logarithmic divergences present in the
$1/R^3$ terms, as mentioned above.},
 $$
\left.{R_\r R_\s\over (uR)^6}(\vp)\right|_{1/\ve}={1\over \ve} \int
ds \int {d^4x\over (ux)^5}\, 2\,\d(x^2-1) x_\r x_\s\, \vp(y)=
{\pi^2\over 4\,\ve} \int ds  \,(5 u_\r u_\s - \eta_{\r\s})\vp(y).
 $$
This means that, as expected, the part that diverges as $\ve\ra 0$
in the distributional sense, is entirely supported on the worldline.
Indeed, we can write the above result as,
 $$
\left.{R_\r R_\s\over (uR)^6}\right|_{1/\ve}= {\pi^2\over 4\,\ve}
\int (5 u_\r u_\s - \eta_{\r\s})\,\d^4(x-y)\,ds.
 $$
Inserting this expression in \eref{polar} we get,
 $$
\left(\pa^\m A^\n_\ve \pa^\a A^\b_\ve\right)_{div}={1\over 64\,\ve}
\int(u^\a u^\m- \eta^{\a\m}) u^\n u^\b\,\d^4(x-y)\,ds.
 $$
Using this result in \eref{thediv} allows one to determine the
divergent part of the energy--momentum tensor as,
 $$
\left.\Theta^{\m\n}_\ve\right|_{div}= {e^2+g^2\over 32\,\ve}
\int\left(u^\m u^\n-{1\over4}\,\eta^{\m\n}\right)\,\d^4(x-y)\,ds.
 $$

This means that we can define a ``renormalized" energy--momentum
tensor for the electromagnetic field as,
 \be \label{trenorm}
T^{\m\n}_{em}={\cal S}^\prime - \lim_{\ve\ra 0}
\left[\Theta^{\m\n}_\ve - {e^2+g^2\over 32\,\ve} \int \left(u^\m
u^\n - {1\over 4} \,\eta^{\m\n}\right)\d^4 (x-y)\,ds\right],
 \ee
where ${\cal S}^\prime - \lim$ means limit in the distributional
sense. What we have shown here is that this limit exists and
represents a well defined distribution. This means in particular
that the four--momentum integrals over an arbitrary finite
three--volume $\int_V d^3x\, T^{\m0}_{em}$ exist, whether or not the
particle at the given instant is inside $V$. If the acceleration of
the particle vanishes sufficiently fast for $t\ra -\infty$, then
also the total four--momentum is finite \cite{rowese}, see eq.
\eref{explicit} for an explicit expression. Notice also that the
expression for $T^{\m\n}_{em}$, apart from being manifestly
Lorentz--invariant, coincides with the naive tensor $\Theta^{\m\n}$
in the complement of the worldline. This feature realizes the
requirement that the energy--momentum tensor should be ``changed
only at the position of the particle". The counterterm $u^\m u^\n$
in \eref{trenorm} can be interpreted as a kind of mass term, while
the term proportional to $\eta^{\m\n}$ is needed to keep
$T^{\m\n}_{em}$ traceless.

\section{Energy--momentum conservation}

Above we have constructed an energy--momentum tensor that gives rise
to finite momentum integrals. Given the (a priori) arbitrariness of
our construction, its physical justification arises from the
fulfillment of local energy--momentum conservation. The check of
this conservation law requires the evaluation of the
four--divergence $\pa_\m T^{\m\n}_{em}$. The present section is
devoted to this evaluation, the result being given in \eref{final}.

We begin by stating the form the (regularized) Maxwell equations
satisfied by $F^{\m\n}_\ve$. From \eref{regfs} and \eref{regh} one
obtains,
 \be
\pa_\m F^{\m\n}_\ve= e \,j^\n_\ve, \quad\quad \pa_\m\widetilde
F^{\m\n}_\ve=g\, j^\n_\ve,\label{regmax}
 \ee
where the regularized unit current,
 \be\label{defcur}
j^\m_\ve\equiv \Box A^\m_\ve,
 \ee
which is still conserved, can be calculated applying one more
derivative to \eref{derpot},
 \be\label{regcur}
j^\m_\ve={\ve^2\over 4\pi}\left( {1\over
(uR)^4}\left[(uR)\,b^\m-(bR)\,u^\m\right]+3(1-(wR))^2{u^\m\over
(uR)^5}+3(1-(wR)) {w^\m\over (uR)^4}\right).
 \ee
The factor of $\ve^2$ arises from the fact that $R^\m R_\m=\ve^2$
and it implies that for $\ve \ra 0$, in the complement of the
wordline $j^\m_\ve$ converges pointwise to zero. More precisely, one
has the distributional limit ${\cal S}^\prime - \lim_{\ve\ra 0}
j^\m_\ve =j^\m$, as implied by \eref{defcur}.

We come now back to the evaluation of $\pa_\m T^{\m\n}_{em}$. Since
the convergence in \eref{trenorm} is in the distributional sense and
since the distributional derivative is a continuous operation, we
can interchange the derivative with the limit,
 \be\label{4div}
\pa_\m T^{\m\n}_{em}= {\cal S}'-\lim_{\ve\ra0} \left(\pa_\m
\Theta^{\m\n}_\ve - {e^2+g^2\over 32\,\ve} \int \left(w^\n - {1\over
4}\,\pa^\n \right)\d^4 (x - y)\,ds\right).
 \ee
Using the regularized Maxwell equations \eref{regmax} one obtains,
 \ba
\pa_\m \Theta^{\m\n}_\ve&=& -\left(e\, F^{\n\m}_\ve +g \,\widetilde
F^{\n\m}_\ve\right)j_{\ve\m}\\
  &=&-\left(e^2+g^2\right)H_\ve^{\n\m}j_{\ve\m} -(e\,f^{\n\m}+g\, \widetilde
  f^{\n\m})j_{\ve\m}.\label{rid}
 \ea
In the terms containing the external field one can simply take the
limit $j_{\ve\m} \ra j_\m$, whereas in the first term one has to
evaluate carefully the distributional limit of the product
$H_\ve^{\n\m}j_{\ve\m}$. Due to the factor $\ve^2$ in front of
\eref{regcur}, this product converges pointwise to zero in the
complement of the worldline; this means that its distributional
limit -- if it exists -- is necessarily supported on the worldline.
On the other hand, the distributional limit ${\cal
S}'-\lim_{\ve\ra0} H_\ve^{\n\m}j_{\ve\m}$ cannot exist. Indeed,
since the four--divergence of a distribution is again a
distribution, we know that the limit in \eref{4div} exists, and this
implies that the divergent contributions of $H_\ve^{\n\m}j_{\ve\m}$
must precisely compensate the $1/\ve$ terms in \eref{4div}. Thus,
actually it is sufficient to  determine the finite contributions of
$H_\ve^{\n\m}j_{\ve\m}$, as $\ve\ra 0$ in the distributional sense.

Using the techniques illustrated in the appendix it can indeed be
shown that one has,
 \be\label{contr}
 H^{\n\m}_{\ve}j_{\ve \m}=\int\left[{1\over
6\pi}\left({dw^\n\over ds}+w^2u^\n\right)-{1\over
32\ve}\left(w^\n-{1\over
4}\,\pa^\n\right)\right]\d^4(x-y)\,ds+o(\ve),
 \ee
where $o(\ve)$ stands for terms that go to zero as $\ve\ra 0$ in the
distributional sense. Using this in \eref{rid} one sees that the
$1/\ve$ terms in \eref{4div} cancel, and one obtains,
 \be\label{final}
\pa_\m T^{\m\n}_{em}=  -\int\left[{e^2+g^2\over
6\pi}\left({dw^\n\over ds}+w^2u^\n\right) +(e\,f^{\n\m}+g\,
\widetilde f^{\n\m})u_\m\right] \d^4(x-y)\,ds.
 \ee
This implies eventually that, when adding the energy--momentum
tensor of the dyon, $T^{\m\n}=T^{\m\n}_{em}+m\int u^\m u^\n
\d^4(x-y)\,ds$, the total energy--momentum tensor is conserved, if
the generalized Lorentz--Dirac equation \eref{lddyon} holds.

If the external field vanishes, $f^{\m\n}=0$, eq. \eref{final} can
be integrated over whole three--space at fixed time $t$, to obtain
the derivative of the total four momentum of the electromagnetic
field, ${dP_{em}^\m\over dt}$. If the four--acceleration of the dyon
vanishes sufficiently fast for $t\ra -\infty$, a further integration
gives then the total four--momentum of the electromagnetic field as,
 \be\label{explicit}
P_{em}^\m(t)=\int d^3x\, T^{0\m}_{em}=-{e^2+g^2\over
6\pi}\left(w^\m(s)+\int_{-\infty}^s
 w^2(\lambda)\,
 u^\m(\lambda)\,d\lambda\right),
 \ee
where $s$ is the proper time of the particle at the instant $t$.

\section{Interpretation and Outlook}

The construction of the energy--momentum tensor performed here
supplies further evidence of the consistency of the classical
dynamics of a radiating dyon system. Indeed, for a system of
particles our construction generalizes simply by replacing in
\eref{trenorm} the counterterm with ${1\over
32\,\ve}\sum_r(e^2_r+g^2_r) \int \left(u^\m_r u^\n_r - {1\over 4}
\,\eta^{\m\n}\right)\d^4 (x-y_r)\,ds_r.$ This subtraction is
sufficient since the mutual interactions do not give rise to
singularities in $\Theta^{\m\n}_\ve$.

A question that arises naturally is whether the energy--momentum
tensor constructed in \eref{trenorm} is determined uniquely. If we
insist on the physical requirement that off the worldline this
tensor should coincide with the original one \eref{naiv}, and if we
enforce duality invariance, a priori the electromagnetic
energy--momentum tensor is indeed determined only modulo a term,
supported on the worldline, of the form,
$$
\Delta T^{\m\n}_{em}=(e^2+g^2) \int h^{\m\n} \d^4(x-y)\,ds,
$$
where $h^{\m\n}$ is a symmetric tensor, of dimension one over
length, constructed with $u^\m$, $w^\m$, $dw^\m/ds$ etc. The
question is now if there exists a tensor $h^{\m\n}$ for which the
modified total energy--momentum tensor $\widehat T^{\m\n}\equiv
T^{\m\n}_{em}+\Delta T^{\m\n}_{em}+m\int u^\m u^\n \d^4(x-y)\,ds$ is
still conserved. Given the above form of $\Delta T^{\m\n}_{em}$ the
answer to this question is the same as in the case of charged
particles, and it has been given in \cite{rowese}: there exists no
$h^{\m\n}\neq 0$ such that $\partial_\m\widehat T^{\m\n}=0$, {\it
whatever modified Lorentz--Dirac equation one imposes on the
particle}. Given the above requirements, and the implicit
assumptions that our dyon is point--like and spinless,
four--momentum conservation fixes therefore \eref{trenorm} uniquely.

Our regularization appears also particularly useful for the
derivation of effective equations of motion for extended objects,
alternative to \cite{kazinski}. In the present case e.g. the self
force can be obtained evaluating $F_\ve^{\m\n}$ in \eref{regfs} at
the worldline $x=y(s)$, using \eref{reghexp}, and then taking $\ve
\ra  0$. The result is,
$$
H_\ve^{\m\n}(y(s))={1\over 8\pi \ve}\left(u^\m w^\n-u^\n w^\m\right)
-{1\over 6\pi} \left(u^\m {dw^\n\over ds}- u^\n {dw^\m\over
ds}\right) +o(\ve).
$$
Using this in the regularized Lorentz equation for dyons,
${dp^\m\over ds}= \left(e F^{\m\n}_\ve+g\widetilde
F^{\m\n}_\ve\right)u_\n$, one sees that the divergent part
renormalizes the mass, and that the finite part amounts to
\eref{lddyon}.

Our regularization scheme admits a simple interpretation in terms of
the retarded Green function $G(x) ={1\over 2\pi} H(x^0) \d(x^2)$ of
the Laplacian $\Box=\pa _\m\pa^\m$. It is indeed immediately seen
that the regularized potential \eref{potreg} is produced by the
regularized Green function $G_\ve(x)={1\over 2\pi} H(x^0)
\d(x^2-\ve^2)$, where $H$ is the Heaviside function, according to
$A_\ve^\m(x)=\int d^4z \,G_\ve(x-z)j^\m(z)$. This regularization
extends naturally to arbitrary dimensions since in even
space--times, $D=2n+4$, the Green--function is $G=
H(x^0)/2\pi^{n+1}(d/dx^2)^n\d(x^2)$, while in odd ones, $D=2n+3$, it
is $G=H(x^0)/2\pi^{n+1}(d/dx^2)^n[H(x^2)/\sqrt{x^2}]$, see
\cite{courant}. The regularized Green function $G_\ve$ in arbitrary
dimensions is then simply obtained operating in $G$ the replacement
$x^2\ra x^2-\ve^2$.

Interpreted in this way our method admits then a natural extension
to extended objects in higher dimensions. For example, for an
electric brane in $D$ dimensions, minimally coupled to a $p$--form
gauge field $B$, we can introduce a retarded regularized potential
-- in Lorentz gauge -- according to $B_\ve(x)=\int d^Dz\,
G_\ve(x-z)j^{(p)}(z)$, where the current $j^{(p)}$ is the
$\d$--function on the brane, i.e. its Poincar\`e dual. This
potential gives rise to a field strength $(p+1)$--form
$F_\ve=dB_\ve$, that is regular on the brane, and hence to the
regularized energy--momentum tensor $\Theta_\ve^{\m\n}={1\over
p!}[(F_\ve F_\ve)^{\m\n}-{1\over 2(p+1)}\,\eta^{\m\n}(F_ \ve F_\ve)]
$. Following the lines of the present paper it should then to be
possible to construct a finite and conserved energy--momentum tensor
for a generic brane -- taking its radiation reaction into account --
providing thus a physical basis for the effective equations of
motion postulated previously, \cite{kazinski}. Moreover, the
knowledge of this tensor should allow also a systematic analysis of
the energy--momentum loss of an extended object, due to the emitted
radiation. We hope to report on this construction soon.

\vskip1truecm

\paragraph{Acknowledgements.}
This work is supported in part by the European Community's Human
Potential Programme under contract MRTN-CT-2004-005104,
``Constituents, Fundamental Forces and Symmetries of the Universe".

\vskip1truecm

\section{Appendix: Evaluation of $H_\ve^{\n\m}j_{\ve\m}$ for
$\ve\ra 0$}

From \eref{regh} and \eref{derpot} one obtains,
 \be\label{reghexp}
H_\ve^{\m\n}={1\over 4\pi(uR)^3}\left[R^\m u^\n+(u^\r w^\n- u^\n
w^\r)R_\r R^\m -(\m\leftrightarrow \n)\right].
 \ee
From this and \eref{regcur} one sees that all terms appearing in the
product $H_\ve^{\n\m}j_{\ve\m}$ are schematically of the form,
 \be\label{serve}
I_\ve\equiv\ve^2 \,{R^{\m_1}\cdots R^{\m_N}\over (uR)^M}
\,\,G(s_\ve),
 \ee
where $G$ is a tensor constructed with $u$, $w$ and $b$, all
evaluated at $s_\ve(x)$, whose tensorial structure we do not
indicate explicitly. One has to apply this expression to a test
function and to consider the limit $\ve\ra 0$. Proceeding as in
\eref{poldiv} one obtains,
 \ba
I_\ve(\vp)&=&  \ve^2 \int ds \int {d^4x\over (ux)^{M-1}}\,
2\,\d(x^2-\ve^2) \,x^{\m_1}\cdots
x^{\m_N}\,G(s) \, \vp(x+y)\label{proto}\\
 &=&  {1\over\ve^{M-N-5}} \int ds \int
{d^4x\over (ux)^{M-1}}\, 2\,\d(x^2-1) \,x^{\m_1}\cdots
x^{\m_N}\,G(s)\left[\vp(y)+\ve \,x^\a \pa_\a\vp(y)+\cdots\right],\nn
 \ea
where we have rescaled $x\ra \ve\, x$, and expanded $\vp(\ve x+y)$
in powers of $\ve$. The values of $M$ and $N$ appearing in
$H_\ve^{\n\m}j_{\ve\m}$ are such that $M-N=4,5,6,7$. Therefore, as
$\ve \ra0$ at most the first three terms of the series above give a
non vanishing contribution. To conclude the evaluation of
$I_\ve(\vp)$ one must eventually perform the integration over
$d^4x$. This integration can be performed by taking multiple
derivatives w.r.t $u^\m$, considered as an independent variable, of
the generating function,
$$
\int d^4x \,{2\d(x^2-1)\over (ux)^n}= {\pi^{3/2}\over (u^2)^{n/2}}
 {\Gamma\left({n\over 2}-1\right)\over \Gamma\left({n+1\over
 2}\right)},
$$
and setting eventually $u^2=1$. The $x$--integration gives thus rise
to polynomials in $u^\m$.

One sees then that as $\ve\ra 0$ $I_\ve(\vp)$ reduces to a (finite)
sum of terms of the kind $1/\ve^l \int
L(s)\,\pa\cdots\pa\,\vp(y)\,ds$, where $l=0,1,2$ and at most two
derivatives on $\vp$ appear. This means that $I_\ve$, as $\ve$ goes
to zero in the distributional sense, is supported on the worldline,
becoming a sum of terms of the type $1/\ve^l \int L(s)\,
\pa\cdots\pa\, \d^4(x-y)\, ds$.
 More precisely, the terms in
$H_\ve^{\n\m}j_{\ve\m}$ with $M-N=4$ converge to zero as $\ve\ra 0$,
the ones with $M-N=5,6$ give rise to finite and simple pole
contributions, while the ones with $M-N=7$ give rise, a priori, also
to double pole contributions. However, by direct inspection one sees
that the double poles cancel. Indeed, from \eref{reghexp} and
\eref{regcur} one sees that the terms in $H_\ve^{\n\m}j_{\ve\m}$
with $M-N=7$, are given by $ I_\ve^7={3 \ve^2\over
(4\pi)^2(uR)^8}R_\m\left(\eta^{\m\n}-u^\m u^\n\right).$ According to
\eref{serve} and \eref{proto} we have $(M=8,N=1)$,
$$
I_\ve^7(\vp)={3\over (4\pi)^2 \ve^2} \int ds \int {d^4x\over
(ux)^7}\, 2\,\d(x^2-1)\,x_\m
  \left(\eta^{\m\n}-u^\m u^\n\right)\left[\vp(y)+\ve \,x^\a \pa_\a\vp(y)+\cdots\right],
$$
and the double pole cancels since $\int {d^4x\over (ux)^7}\,
2\,\d(x^2-1)\,x_\m= {8\pi\over 15}\,u_\m$. Only finite and simple
pole terms survive then in $H_\ve^{\n\m}j_{\ve\m}$, and a
straightforward but a bit lengthy calculation gives \eref{contr}.

\end{document}